\documentclass[journal, 12pt, onecolumn]{IEEEtran}

\usepackage{cite}
\usepackage{xcolor,soul,framed} 
\usepackage{amsmath, amssymb}
\allowdisplaybreaks
\colorlet{shadecolor}{yellow}
\usepackage[pdftex]{graphicx}
\graphicspath{{../pdf/}{../jpeg/}}
\DeclareGraphicsExtensions{.pdf,.jpeg,.png}

\usepackage{amsmath}

\usepackage{verbatim}

\usepackage{array}
\usepackage{mdwmath}
\usepackage{mdwtab}
\usepackage{eqparbox}
\usepackage{url}

\usepackage[caption=false,font=footnotesize]{subfig} 

\begin{document}

\pagestyle{empty}      
\thispagestyle{empty}  

\bstctlcite{IEEEexample:BSTcontrol}
    \title{Evaluation of EMF Exposure to Throughput Ratio for Sustainable 5G Networks}
  \author{Dinh~Long~Trinh, Shanshan~Wang, Joe~Wiart \\
      Chaire C2M, LTCI, Télécom Paris, Institut Polytechnique de Paris, 91120 Palaiseau, France;\\
      e-mail: \{dinh.trinh, shanshan.wang, joe.wiart\}@telecom-paris.fr
}

\markboth{
}{Roberg \MakeLowercase{\textit{et al.}}: Evaluation of EMF Exposure in 5G EN-DC Networks}

\maketitle
\thispagestyle{empty}   
\pagestyle{empty}


\begin{abstract}
This paper builds stochastic geometry frameworks for analyzing downlink electromagnetic field (EMF) exposure and efficiency in 5G multi-connectivity networks, using 5G E-UTRAN New Radio - Dual Connectivity (EN-DC) configuration as a representative use case. The Poisson point process (PPP) and the $\beta$-Ginibre point process ($\beta$-GPP) are used to model the spatial distribution of base stations (BSs), where $\beta$-GPP effectively captures the repulsion observed in real deployments. We derive tractable expressions for the distribution of EMF exposure and validate the framework through both Monte Carlo simulations and real BS data from Paris. In addition to conventional metrics, we introduce the \textit{Radiated Energy per Bit Transmitted in the Downlink} (REBT-DL), which accounts for throughput and received power. Results show that network configuration significantly affect exposure and REBT-DL, highlighting the relevance of energy-aware deployment strategies and confirming the proposed approach as a comprehensive tool for sustainable network evaluation. The results also confirm that $\beta$-GPP provides a more accurate fit to practical deployments than PPP.
\end{abstract}

\begin{IEEEkeywords}
Stochastic geometry, $\beta$-Ginibre point process, EMF exposure, sustainability, REBT-DL, 5G EN-DC
\end{IEEEkeywords}

%
\IEEEpeerreviewmaketitle


\section{Introduction}\label{sec:introduction}


\IEEEPARstart{T}{he} rapid growth of mobile subscribers and data traffic drives the densification of base stations (BSs) as a promising solution \cite{andrews2014will}. Among various network performance metrics, sustainability, which seeks to balance network performance with environmental responsibility, has emerged as a key consideration in future network design. 
Energy efficient architectures, renewable powered operation, and Artificial intelligence (AI) driven power management can reduce carbon emissions and costs, enabling greener networks, yet public concerns about electromagnetic field (EMF) exposure and health effects persist.
Therefore, deploying new BSs must be carefully managed to ensure compliance with EMF limits.

EMF exposure in wireless networks is typically assessed through in-situ measurements, e.g., drive tests, and sensor networks. Numerical methods, such as ray-tracing or ray-launching \cite{celaya20202g}, provide accurate results but are computationally intensive for large-scale networks. Hence, stochastic approaches are often preferred for deriving statistical insights such as the probability of exceeding regulatory limits.

Stochastic Geometry (SG) provides a tractable framework for network-level analysis by modeling BSs and users as random point processes, from which analytical expressions for metrics like coverage, throughput, and EMF exposure can be derived. Among the commonly used point processes, the Poisson point process (PPP) is widely adopted for its tractability, but its assumption of spatial independence among BSs fails to reflect the regularity observed in real deployments. To overcome this limitation, various non-Poisson point processes have been employed to better model real-world BS deployments. Among these, the $\beta$-Ginibre point process ($\beta$-GPP), originated from GPP, has been proposed to capture the spatial repulsion between BSs \cite{deng2014ginibre}. The parameter $\beta$ $(0 < \beta \leq 1)$ controls the degree of repulsion, where $\beta = 1$ corresponds to the standard GPP, and $\beta \to 0$ leads to convergence toward PPP.

Currently, wireless networks are evolving into heterogeneous, multi-RAT 5G architectures to enable a smooth and sustainable transition toward 5G-Advanced and beyond. The coexistence between different radio access technologies (RAT) introduce new opportunities for infrastructure sharing and resource optimization. A practical and widely deployed realization of such multi-connectivity is the 5G E-UTRAN New Radio - Dual Connectivity (EN-DC) configuration, which enables simultaneous 4G and 5G downlink transmission. However, most existing studies based on SG focus primarily on single-tier networks \cite{deng2014ginibre, gontier2024joint, al2020statistical, wang2019meta}, leaving a gap in the modeling of EMF exposure and network performance under multi-connectivity scenarios. In particular, simultaneous transmissions from multiple layers of networks necessitate a joint assessment of exposure and throughput. 

To this end, we consider the Radiated Energy per Bit Transmitted in the Downlink (REBT-DL), a metric quantifying the ratio between total EMF exposure and throughput \cite{liu2024assessment}. Focusing on REBT-DL provides a user-centric view of EMF exposure, linking it with throughput performance and complementing conventional power-based evaluations for sustainable network design.

The main contributions of this paper include:
\textbf{\itshape(i)} Deriving the Cumulative Distribution Function (CDF) for downlink EMF exposure in 4G, 5G, and EN-DC networks, considering both PPP and $\beta$-GPP spatial models;
\textbf{\itshape(ii)} Analyzing the REBT-DL metric, capturing the trade-off between EMF exposure and throughput from a sustainability perspective;
\textbf{\itshape(iii)} Fitting $\beta$-GPP with real BS location data from the Cartoradio database, examining Orange’s network in Paris with 5G BSs at 3500 MHz and 4G BSs at 2600 MHz.

\section{System Model}\label{sec:system_model}
\subsection{Network and Spatial Modeling}
A downlink multi-RAT cellular network is considered, where 4G LTE and 5G NR BSs are modeled as independent spatial point processes $\Psi_{\text{4G}}$ and $\Psi_{\text{5G}}$ of densities $\lambda_{\text{4G}}$ and $\lambda_{\text{5G}}$, respectively. Each process can follow either a PPP or a $\beta$-GPP distribution.
For a PPP, the probability density function (PDF) of the distance $r$ between the typical user and its serving BS is given by \cite{baccelli2024random}: $f_R(r) = 2 \pi \lambda r e^{-\pi \lambda r^2}$.

For the $\beta$-GPP with $0 < \beta \leq 1$, the squared distances ${|X_i|^2}$ follow a thinned sequence of independent random variables $Q_k \sim \text{Gamma}(k, c / \beta)$. Defining $v = \frac{\beta}{c} Q_k$, the normalized distances $\frac{\beta}{c}|X_i|^2$ are distributed as $\Gamma(k,1)$ with PDF:
\begin{align}
f_k(v) = v^{k-1} e^{-v} \, / \, \Gamma(k)
\label{eq:pdf:gpp}
\end{align}
Let $\{\epsilon_k\}$ be Bernoulli$(\beta)$ indicators for whether $Q_k$ is kept ($\epsilon_k=1$) or removed ($\epsilon_k=0$). The serving BS $X_0$ is then $X_s$ if $\epsilon_s=1$, while all other BSs are either removed or located farther than $X_s$, $\{X_0 = X_s\} = \{\epsilon_s = 1\} \cap \mathcal{A}_s,$ where $\mathcal{A}_s = \{\epsilon_k = 1, |X_k| > |X_s|\} \cup \{\epsilon_k = 0\} \quad \forall k \in \mathbb{N} \setminus \{s\}.$

We assume a fully loaded network with $\lambda_{UE} \gg \lambda_{BS}$, and a closest BS association policy.

This general multi-RAT framework can represent either standalone 4G/5G networks or EN-DC configuration.

\subsection{Propagation Model}
The large-scale path loss follows a bounded model:
\begin{align}
\ell(r^2) =
1 \cdot \mathbf{1}_{\{r < D\}} + \left(r^2 \right)^{-\alpha / 2} \cdot \mathbf{1}_{\{r \geq D\}}
\label{eq:path:loss}
\end{align}
where $\alpha$ is the path-loss exponent and $D$ is the guard-zone radius. The notation $\mathbf{1}_{\{\cdot\}}$ denotes the indicator function. Small-scale fading is modeled as Rayleigh fading with unit mean.

5G BSs employ dynamic beamforming (BF) with three 120° antenna arrays. The antenna gain pattern is modeled as:
\begin{align}
G_{k, {\text{5G}}} = 
G_{\text{5G}}^{\text{max}} \cdot \mathbf{1}_{\{|\theta|\leq \omega\}}
\label{eq:antenna:gain}
\end{align}
where $\omega \in [0, 2\pi/3]$ is the beamwidth. The orientations of interfering BS antennas are uniformly distributed, thus $G_{k, {\text{5G}}}$ can be treated as a Bernoulli random variable with mean $\eta G_{\text{5G}}^{\text{max}}$, where $\eta = \frac{3\omega}{2\pi}$ is the beam alignment probability.
4G BSs are modeled as omnidirectional transmitters with constant gain $G_{\text{4G}}$.

We assume an interference-limited network. For each RAT $l \in \{\text{4G}, \text{5G}\}$, the signal-to-interference ratio (SIR) at the typical user is defined as $\text{SIR}_l = {S_l}/{I_l}$, where the signal $S_l$ and interference $I_l$ are given by:

\begin{equation}
\begin{aligned}
&S_l = P_l \left|h_{s, l}\right|^2 \, \ell\left(r_{s, l}^2\right) \\
&I_l = \sum_{X_{k, l} \in \Psi_l \setminus {X_{s, l}}}
P_l G_{k, l} \left|h_{k, l}\right|^2  \ell\left(r_{k,l}^2\right)
\end{aligned}
\end{equation}

For compact notation, the effective transmit powers are defined as $P_5 = P_5 G^{\text{max}}_{\text{5G}}$ and $P_4 = P_4 G_{\text{4G}}$. In the following equations, the subscripts “4G” and “5G” are abbreviated as “4” and “5” for simplicity.

\subsection{EMF Exposure Metric}
The total downlink EMF exposure at the typical user combines contributions from both RATs:
\begin{align}
\mathcal{E} = \sum_{l \in \{4G,5G\}} S_{l} + I_{l}
\label{eq:EMFE:def}
\end{align}
representing the received electromagnetic power from all links.

\subsection{REBT-DL}
REBT-DL captures the energy and environmental efficiency of the network, quantifying the amount of radiated power needed per transmitted bit. REBT-DL is defined as:
\begin{align}
Y \!=\! \frac{\text{EMF exposure}}{\text{Throughput}} \!=\! \frac{\sum_{l \in \{4G, 5G\}} S_{l} + I_{l}}{\sum_{l \in \{4G, 5G\}} \! W_{l}  \log_2(1 \!+\! \text{SIR}_{l})}
\end{align}
Here, $W_{l}$ denotes the bandwidth of RAT $l$.

\section{Mathematical Framework}\label{sec:framework}
\subsection{CDF of EMF Exposure}
Given the definition of total downlink EMF exposure in \eqref{eq:EMFE:def}, we derive the CDF of $\mathcal{E}$ using Gil–Pelaez inversion theorem \cite{gil1951note}, denoted as:
\begin{align}
F_{\mathcal{E}}(\tau)
&= \frac{1}{2} - \int_0^{\infty} \frac{\mathrm{Im} \left\{ \Phi_{\mathcal{E}}(t) \, e^{-j t \tau} \right\}}{\pi t} \, dt
\label{eq:CDF:Gil:Palaez}
\end{align}
\begin{align}
f_{\mathcal{E}}(\tau)
&= \int_0^{\infty} \frac{\mathrm{Re} \left\{ \Phi_{\mathcal{E}}(t) \, e^{-j t \tau} \right\} }{\pi} \, dt
\label{eq:CDF:PDF:Gil:Palaez}
\end{align}

where $\Phi_{\mathcal{E}}(t)$ is the characteristic function (CF).

\noindent\textbf{Proposition 1.}
The CF of the EMF exposure under a single-tier 5G network is:
\begin{align}
\Phi_{\mathcal{E}_5}(t) 
= \int_0^\infty \frac{\Phi_{I_5}(t \mid \theta_5 )}{1 - jt P_5 \ell(\theta_5)} \, d\Pi_5(\theta_5)
\label{eq:CF:EMFE:PPP:single:BF}
\end{align}
\noindent {\itshape For PPP} $\left(\theta_5 = r_5; \, d\Pi_5 (\theta_5) = f_R(r_5) \, dr_5 \right)$:
{\small
\begin{align}
\Phi_{I_5}\!\left( t \mid r_5\right)
&= \mathbf{1}_{\{r_5 \geq D\}} \, \exp\left(\displaystyle \frac{j t \eta P_5 \, 2\pi\lambda_5 \, r_5^{2 - \alpha}}{\alpha - 2}  \, \Omega\left(\frac{jtP_5}{r_5^{\alpha}} \right) \right) \nonumber \\
& + \ \mathbf{1}_{\{r_5 < D\}} \, \exp\left( \displaystyle \frac{jt \eta P_5 \,\pi\lambda_5 \left(D^2 - r_5^2\right)}{1 - jt P_5} \right) \nonumber \\
&\quad \times \exp\left(\displaystyle \frac{j t \eta P_5 \, 2\pi\lambda_5 D^{2 - \alpha}}{\alpha - 2} \, \Omega\left(\frac{jtP_5}{D^{\alpha}} \right) \right)
\label{eq:CF:interference:PPP:single:BF}
\\
\text{where }& \; \Omega(x) = {}_2F_1\left(1, 1 - \frac{2}{\alpha}; 2 - \frac{2}{\alpha}; x\right) \nonumber 
\end{align}
}

\noindent {\itshape For $\beta$-GPP} $\left(\theta_5 = \frac{c_5}{\beta_5}r_5^2 = u; \, d\Pi_5(\theta_5) = \Upsilon(u) \, du \right)$:
\begin{align*}
\Upsilon(u) = \beta_5\sum_{s=1}^\infty f_s(u) \prod_{k \neq s} \left( 1 - \beta_5 + \beta_5 \, \frac{\Gamma(k, u)}{\Gamma(k)} \right)
\end{align*}
\begin{align}
\Phi_{I_5} \!\left( t \! \mid \! u \right)=\prod_{k \neq s} \! \left( \! 1 \!-\! \eta \beta_5 \!+\! \! \int_{u}^{\infty} \! \frac{\eta \beta_5 f_k(v)}{ 1 \!- \! jt P_5 \,\ell\left(\!\frac{\beta_5}{c_5}v \! \right) }  dv \! \right)
\label{eq:CF:interference:GPP:single:BF}
\end{align}
\noindent{\itshape Proof.} The proof is provided in Appendix~\ref{Appendix:A} and Appendix~\ref{Appendix:B}.

In single-tier 4G networks, the CF of the EMF exposure can be simplified by setting $\eta = 1$, and the terms $S$ and $I$ no longer need to be treated separately. Under PPP deployment, the CF of the EMF exposure is given by:
\begin{align*}
\Phi_{\mathcal{E}_4}(t) \!=\!  \exp \! \left( \! \frac{jt P_4 2 \pi \lambda_4 D^{2 - \alpha}}{\alpha - 2} \! \Omega\left( \! \frac{jt P_4}{D^\alpha}\! \right) \!\right) \! \exp \! \left( \! \frac{jt P_4  \pi\lambda_4 D^2}{1 \!-\! jt P_4} \!\right) 
\end{align*}

\noindent and under $\beta$-GPP deployment:
\begin{align}
\Phi_{\mathcal{E}_4}(t) &= \prod_{k=1}^{\infty} \! \left(1  -  \beta_4  +  \int_0^{\infty} \! \frac{\beta_4 f_k(v)}{1 \! - \! j t P_4 \ell\left(\frac{\beta_4}{c_4} v \right)} \, dv \right) \nonumber
\end{align}

\begin{figure}
  \begin{center}
  \includegraphics[trim={1cm 0.6cm 1cm 1cm}, clip, width=0.75\textwidth]{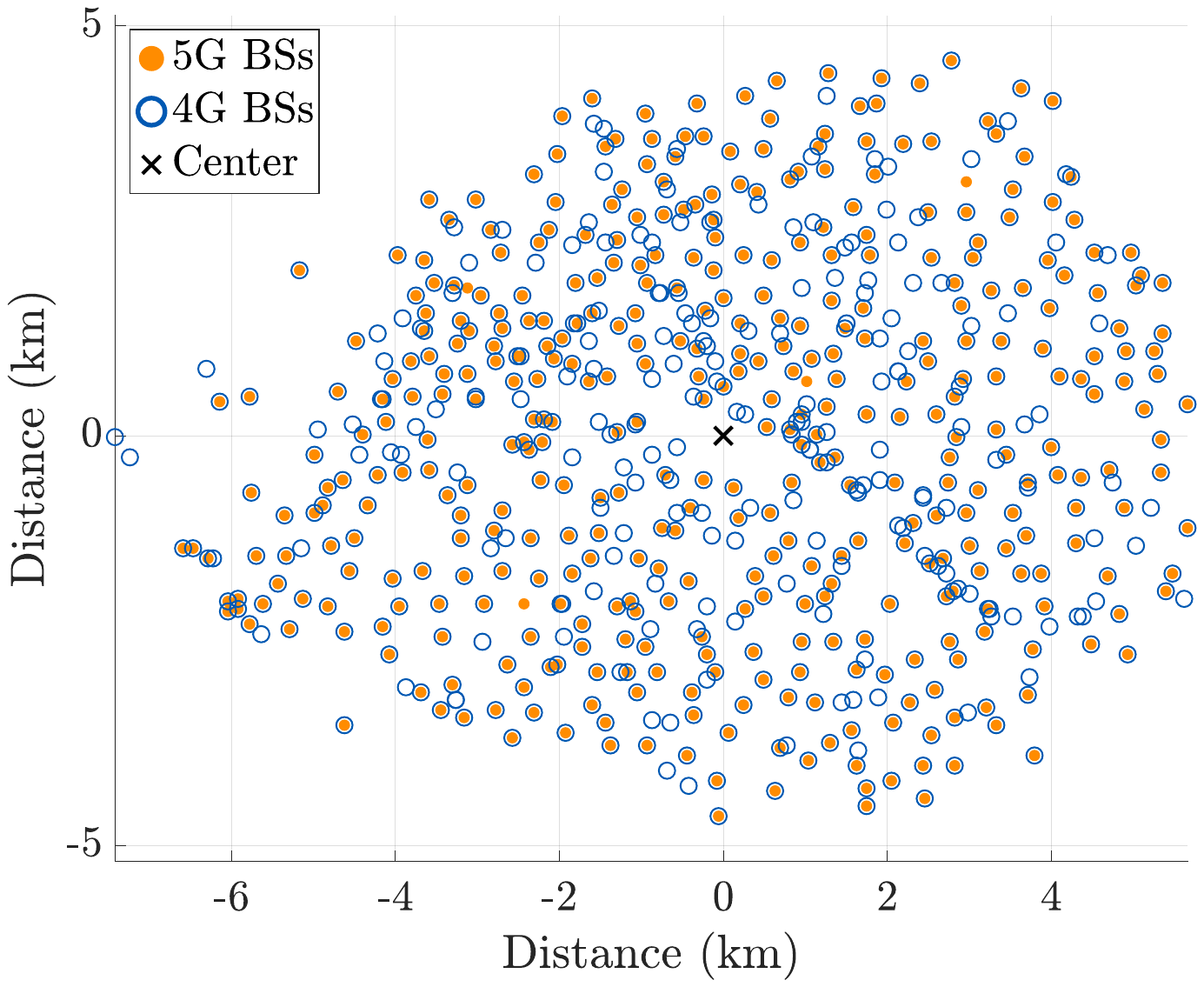}
  \caption{Relative locations of Orange 4G (2600 MHz) and 5G (3500 MHz) BS in Paris under Lambert projection.}\label{fig:1}
  \end{center}
\end{figure}

In commercial EN-DC deployments, 5G is typically co-located with existing 4G sites to take advantage of the established LTE coverage and infrastructure. Indeed, based on data from \cite{ANFR_Cartoradio} (Fig.~\ref{fig:1}), nearly all 5G BSs are co-located with 4G sites. For analytical tractability, each 4G BS is assumed to have a co-located 5G BS with probability $p$, indicated by $\zeta \sim \mathrm{Bernoulli}(p)$, and standalone 5G BSs are ignored.

\noindent\textbf{Proposition 2.}
The CF of the total EMF exposure in the EN-DC networks is:
\begin{align}
\Phi_{\mathcal{E}^{\star}}(t)
&= \int_0^\infty \Phi_{S^{\star}}(t \mid \theta_4) \; \Phi_{I^{\star}}(t \mid \theta_4) \, d\Pi_4(\theta_4)
\label{eq:CF:EMFE:PPP:ENDC:BF}
\end{align}
where $\Phi_{S^{\star}}(t \mid \theta_4) = \frac{1}{1 \!-\! jtP_4 \ell(\theta_4)} \! \left( 1-p + \frac{p}{1 \!-\! jt P_5 \ell \left(\theta_4\right)} \right)$

\noindent {\itshape For PPP:}
\begin{align}
\Phi_{I^{\star}}(t\mid r_4) 
&= \exp \Bigg( -2 \pi \lambda_4 \int_{r_4}^\infty \Bigg( 1 - \frac{1}{ 1 - jt P_4 \ell\left( r^2 \right)} \nonumber \\
& \qquad \left( 1 - p \eta + \frac{p \eta}{1 - jt P_5 \ell\left( r^2 \right) } \right) \Bigg) r \, dr \Bigg)
\end{align}

\noindent {\itshape For $\beta$-GPP:}
\begin{align}
\Phi_{I^{\star}}(t\mid u) 
&= \prod_{k \neq s} \Bigg( 1 - \beta_4 + \int_u^\infty \frac{\beta_4}{1 - jt P_4 \ell\left( \frac{\beta_4}{c_4}v \right)} \nonumber\\
&\quad \left( 1\! - \!p\eta\! +\! \frac{p\eta}{1 \!-\! jt P_5 \ell\left( \frac{\beta_4}{c_4} v \right)} \right) f_k(v) \, dv \Bigg)
\end{align}
{\itshape Proof.} The proof is provided in Appendix~\ref{Appendix:C}.

\subsection{CDF of REBT-DL}

\noindent\textbf{Theorem 1.}
The CDF of the REBT-DL under a single-tier 5G network is:
\begin{align}
F_Y(y) &= \int_0^\infty \!\!\!\! \int_0^\infty \! \! \left( \frac{1}{2} \!-\! \int_{0}^{\infty} \! \frac{\mathrm{Im} \left\{\Phi_{I_5} \!(t \!\mid \! \theta_5 ) \, e^{-jt {I}_5^\star}\right\}}{\pi t} \, dt \right) \nonumber \\
& \qquad \qquad \quad e^{-h} \, dh \, d\Pi_5(\theta_5)
\label{eq:CDF:REBT:5G:PPP}
\end{align}
where ${I}_5^\star$ is the unique solution to $g(I_5; s_5) = 0$, with $s_5 = h P_5 \ell(\theta_5)$, and the function $g(\cdot; \cdot)$ for each RAT $l \in \{\text{4G,5G}\} $ is defined as:
\begin{align}
g(i_l;s_l) = s_l + i_l - y \, W_l \, \log_2 \left(1 + s_l/i_l\right)
\end{align}
{\itshape Proof.}
The proof is given in Appendix~\ref{Appendix:D}.

For the single-tier 4G case, we follow the 5G analysis and omit dynamic BF.

\noindent\textbf{Theorem 2.}
The CDF of REBT-DL under 5G EN-DC network is given by:
\begin{align}
F_{Y^{\star}}(y) \! &= \! \iiint \!\!\! \int_0^\infty \! \left(p\,F_{I_5}\left(I_{5, co}^\star\right) + (1-p)\,F_{I_5}(I_{5, no-co}^\star) \right) \nonumber\\
& \quad \quad \quad \quad f_{I_4}(i_4) \, e^{-h_4} \, e^{-h_5} \,d_{i_4} \, dh_4 \, dh_5 \, d\Pi_4(\theta_4) 
\label{eq:CDF:REBT:ENDC}
\end{align}
where $ I_{5, co}^\star $ is the unique solution of $ g(I_5; s_5) + g(i_4; s_4) = 0 $, with $s_4 = P_4 h_4 \ell(\theta_4)$ and $s_5 = P_5 h_5 \ell(\theta_5)$. The variable $\theta_5$ denotes the 5G serving distance, equal to $\theta_4$ when the 4G and 5G serving BSs are co-located, and $\theta_5>\theta_4$ otherwise.

The CDF of the interference $F_{I_5}(\cdot)$ is evaluated analogously to Proposition 2 but without the term $(1-jtP_4\ell(\cdot))^{-1}$ in the CF. 
In the non-co-located mode, it is averaged over the conditional 5G serving distance as
\begin{align} 
F_{I_5}(x)
= \int_{\theta_4}^{\infty} F_{I_5}(x \mid \theta_5) \, f_{\Theta_5|\Theta_4}(\theta_5|\theta_4) \,d\theta_5
\end{align}
where for the PPP:
\begin{align}
f_{R_5|R_4}(r_5|r_4)
= 2\pi p\lambda_4 r_5 \exp\left(-\pi p\lambda_4(r_5^2-r_4^2)\right)
\end{align}
and for the $\beta$-GPP:
\begin{align}
f_{V|U}(v|u)
&=\sum_{n\ne s} 
\frac{\beta_4 p f_n(v)}{1-\beta_4+\beta_4 \left((1-p)\frac{\Gamma(n,u)}{\Gamma(n)}+p\frac{\Gamma(n,v)}{\Gamma(n)}\right)} \nonumber \\
& \prod_{k\ne s} 
\frac{1 - \beta_4 + \beta_4 \left((1-p)\frac{\Gamma(k,u)}{\Gamma(k)}+p\frac{\Gamma(k,v)}{\Gamma(k)}\right)}{1 - \beta_4 + \beta_4\frac{\Gamma(k,u)}{\Gamma(k)}} 
\end{align}
The 4G interference PDF $f_{I_4}(i_4)$ is obtained from the single-tier CF in Proposition 1.

\noindent {\itshape Proof.}
The proof is given in Appendix~\ref{Appendix:E}.

\begin{table}
\centering
\caption{System Parameters}
\label{tab:details}
\renewcommand{\arraystretch}{1.1}
\begin{tabular}{|c|c||c|c|}
\hline
$P_5$ (dBm) & 51 & $P_4$ (dBm) & 45 \\
$\lambda_5$ ($m^{-2}$) & 5.1244e-6 & $\lambda_4$ ($m^{-2}$) & 7.5294e-6  \\
$\beta_5$ & 0.83 & $\beta_4$ & 0.75 \\
$W_5$ (MHz) & 90 & $W_4$ (MHz) & 20 \\ $D$ (m) & 40 & $\eta$ & 0.0469 \\
$p$ & 0.7 & $\alpha$ & 4 \\
\hline
\end{tabular}
\end{table}

\section{Numerical Results}\label{sec:results}
In this section, we present numerical results for the CDF of EMF exposure and REBT-DL, and validate the framework through Monte Carlo simulations and real datasets in Paris. We also fit the actual BS locations to the $\beta$-GPP and analyze the analytical framework described in Section~\ref{sec:system_model}. To avoid boundary effects, we consider only the BSs located inside a disk centered at coordinates $(48.86^\circ \text{N}, 2.33^\circ \text{E})$ with radius $R = 4500 \text{m}$ as shown in Figure \ref{fig:1}.
By fitting $J$-function proposed in \cite{gontier2024modeling}, we obtain $\beta_4 = 0.75$ and $\beta_5 = 0.83$, as the best fits for empirical 4G and 5G networks respectively. In the EN-DC mode, $\beta_4$ remains unchanged, and based on Fig.~\ref{fig:1}, we set the co-location probability of 4G and 5G BS to $p = 0.7$. The half-power beamwidth of the antenna radiation pattern is 0.0982 rad, which gives $\eta = 0.0469$. The system parameters extracted from the operator’s database \cite{ANFR_Cartoradio} are given in Table~\ref{tab:details}.

It can be observed that the $\beta$-GPP model closely matches the experimental data, while the PPP model tends to underestimate exposure. This confirms that incorporating spatial repulsion between BS leads to a more realistic representation of real-world deployments.

\begin{figure}[ht]
  \begin{center}
  \includegraphics[trim={1cm 1.3cm 1cm 1.5cm}, clip, width=0.75\textwidth]{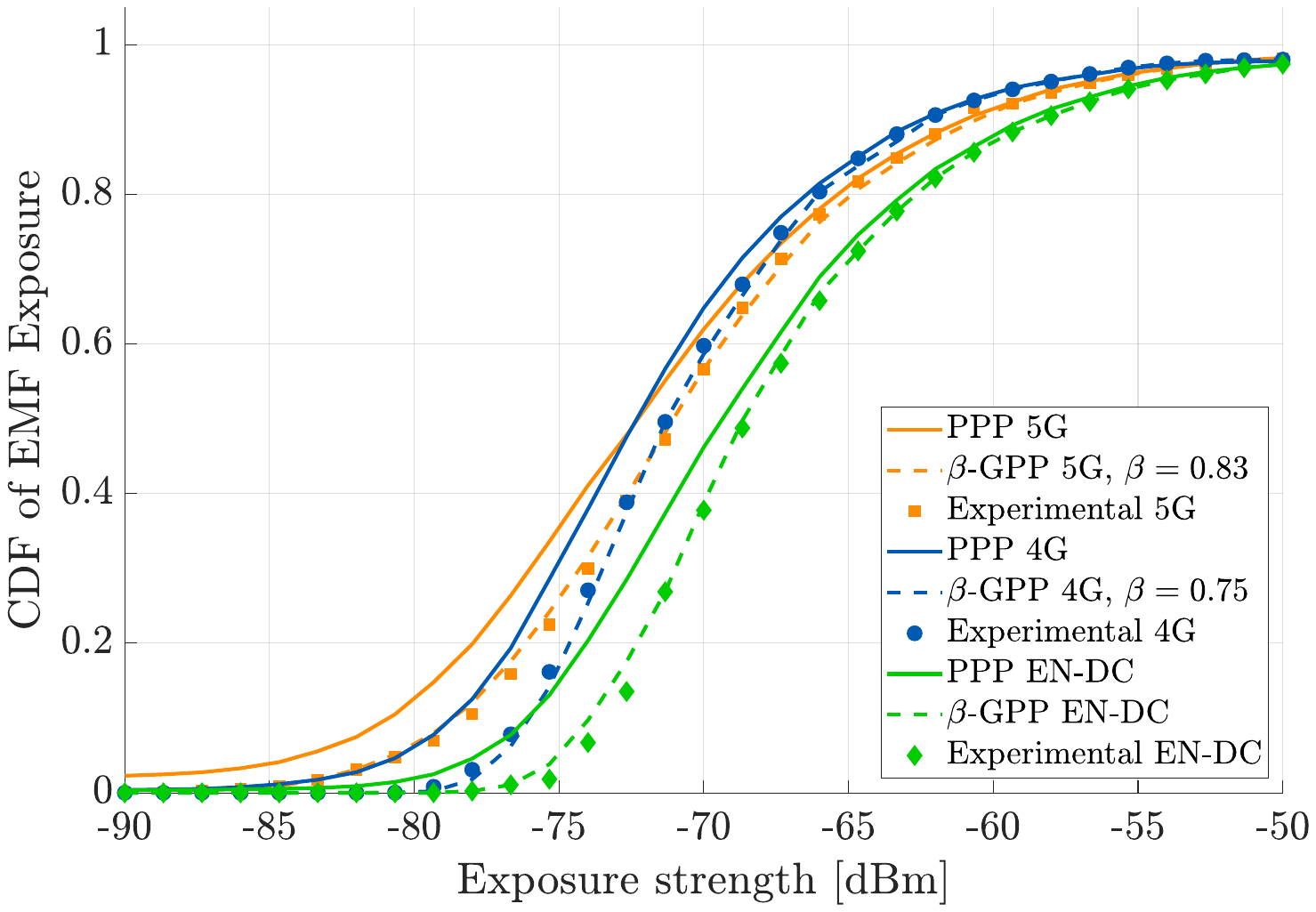}
  \caption{CDF of EMF exposure under PPP, $\beta$-GPP, and Experimental.}\label{fig:2}
  \end{center}
\end{figure}

In Fig.~\ref{fig:2}, the 5G network exhibits slightly higher exposure than 4G due to dynamic beamforming, which concentrates radiated energy within narrower main lobes. It can also be observed that at low exposure thresholds, the 4G curve lies slightly to the right of 5G, indicating that more users experience higher EMF exposure in 4G. This is mainly because 4G employs omnidirectional antennas that radiate uniformly in all directions, resulting in stronger interference and higher non-user exposure levels. However, at higher thresholds, the 5G curve surpasses that of 4G, reflecting that beamforming can generate higher instantaneous received power within the main lobe. While 5G generally reduces average exposure due to directional transmission and lower interference, users located under favorable alignment conditions may experience stronger localized exposure.

Under the EN-DC configuration, where both 4G and 5G downlink transmissions are active, the rightward shift of the curve becomes more pronounced, indicating the accumulative exposure from two coordinated RATs and {thus higher aggregate EMF levels compared to single-tier scenarios.} Another observation is that, in both the REBT-DL and EMF exposure CDF plots, the PPP curve consistently lies above the $\beta$-GPP curve. This implies that, for a given exposure threshold, a larger fraction of users in the PPP model experience lower received power than in the $\beta$-GPP model, caused by the repulsion displayed between BS locations. 
As noted in \cite{andrews2011tractable}, the coverage probability derived under PPP acts as a lower bound for realistic deployments; thus, PPP networks yield lower throughput than $\beta$-GPP ones, resulting in a higher REBT-DL ratio. This confirms that the $\beta$-GPP model not only better fits spatial distributions but also captures realistic exposure–throughput trade-offs.

\begin{figure}
  \begin{center}
  \includegraphics[trim={1cm 1.3cm 1cm 1.5cm}, clip, width=0.75\textwidth]{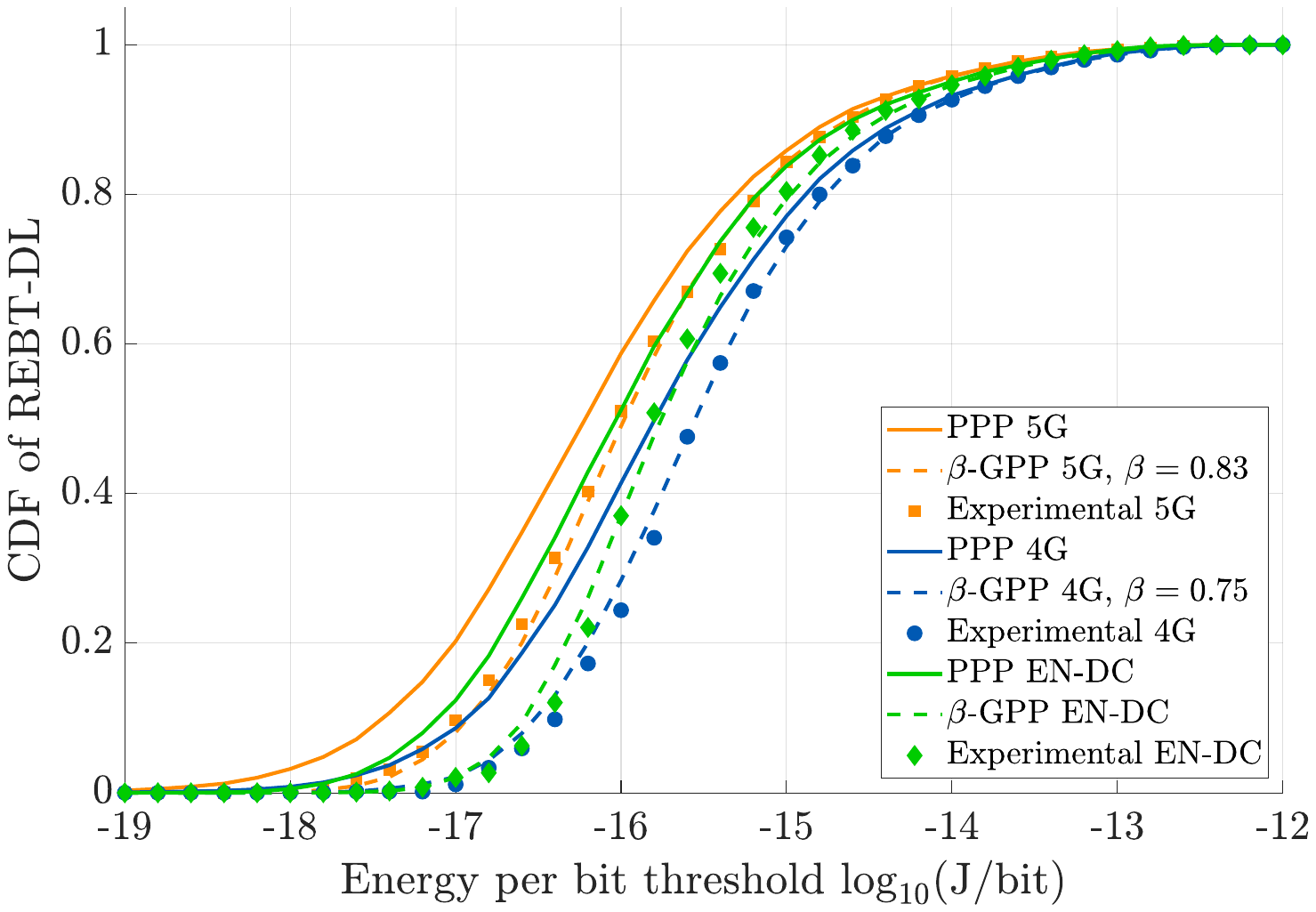}
  \caption{CDF of REBT-DL under PPP, $\beta$-GPP, and Experimental.}\label{fig:3}
  \end{center}
\end{figure}

As shown in Fig.~\ref{fig:3}, the single-tier 5G configuration consistently outperforms both EN-DC and 4G in terms of REBT-DL, demonstrating superior energy efficiency and sustainability. While EN-DC exhibits the highest EMF exposure due to simultaneous 4G and 5G transmissions, its dual connectivity significantly enhances data throughput, which partially compensates for the increased exposure and results in a lower REBT-DL compared to 4G. In contrast, the single-tier 5G network benefits from optimized NR operation and reduced signaling overhead, achieving the best balance between radiated energy and throughput. These results highlight the importance of energy-aware deployment strategies and validate EN-DC as a transitional yet sustainable solution toward fully standalone 5G networks.

\section{Conclusion}\label{sec:conclusion}
This paper studied downlink EMF exposure in 5G EN-DC networks using PPP and $\beta$-GPP models. We derived tractable integral forms for the distribution of EMF exposure, and validation with Orange’s 4G (2600 MHz) and 5G (3500 MHz) deployments in Paris confirmed that the $\beta$-GPP provides a more accurate and flexible spatial model than PPP. We further introduced the REBT-DL metric, which jointly accounts for EMF exposure and throughput, offering a more user-centric and sustainable assessment than average-power metrics. Future work will extend the framework to advanced beamforming, blockages, uplink, heterogeneous architectures, and inter-operator exposure analysis.

\section*{Acknowledgment}
This work benefited from a government grant managed by ANR agency under the France 2030 program, reference ANR-22-PEFT-0008.


%

\appendices
\section{Proof of the CF of EMF exposure in a 5G PPP} \label{Appendix:A}
The CF of received power is defined as $\Phi_{\mathcal{E}_5}(t) = \mathbb{E}_{r_{s}} \left[ \Phi_{S_5} \left(t \mid r_{s}^2 \right) \; \Phi_{I_5} \left( t \mid r_{s}^2 \right) \right]$, where $\Phi_{S_5} \left(t \mid r_{s}^2 \right)$ and $\Phi_{I_5} \left( t \mid r_{s}^2 \right)$ are given by:

\begin{align}
\Phi_{S_5} \! \! \left( t \! \mid \! r_s^2 \right) &\overset{(a)}{=} \frac{1}{1 - jt P_{5} \ell(r_{s}^2)} \\
\Phi_{I_5} \! \! \left( t \! \mid \! r_s^2 \right) 
&= \mathbb{E}_{\Psi_5} \left[ \prod_{X_k \neq {X_s}} \mathbb{E}_{G} \!\left[\frac{1}{1 - jt P_{5} G_k \ell\left(r_{k}^2\right)} \right] \mid r_{s}^2 \right] \nonumber \\
&\overset{(b)}{=} \! \exp\!\left(\! -2\pi\lambda_5 \! \! \int_{r_{s}}^\infty \! \! \left( \! \eta \!-\! \frac{\eta}{1 \!-\! jt P_{5} \ell(r^2)} \! \right) \! r \, dr \! \! \right)
\label{eq:proof:appendix:A:3}
\end{align}
where (a) is obtained by taking the expectation over $|h|^2$, and (b) is obtained by taking the expectations over $|h|^2$ and $G$, followed by applying the Probability Generating Functional (PGFL) of a PPP.

\noindent{\textbf{For $\boldsymbol{r_{s}} < \boldsymbol{D}$}, \eqref{eq:proof:appendix:A:3} becomes:}
\begin{align}
\Phi_{I_5}\left( t \mid r_{s} \right) \nonumber &= \exp\left( \frac{jt \eta P_{5} \, \pi\lambda_5 \left(D^2 - r_s^2 \right)}{1 - jt P_{5}} \right) \nonumber\\
&\times\exp\left( 2\pi\lambda_5 \int_D^\infty \frac{jt \eta P_{5}}{r^\alpha - jt P_{5}} \, r \, dr \right)
\label{eq:proof:appendix:A:4}
\end{align}

\noindent Let $\displaystyle u = \frac{D}{r} \Rightarrow r = \frac{D}{u}, dr = -\frac{D}{u^2} \, du$. The integrand becomes:
\begin{align}
\int_D^{\infty} \frac{1}{r^\alpha - j t P_{5}} \, r \, dr \nonumber &= \int_1^0 \frac{ \, D \, u^{\alpha - 1}}{D^\alpha - j t P_{5} \, u^\alpha} \cdot \left(-\frac{D}{u^2}\right) \, du \nonumber \\
&= \frac{D^{2 - \alpha}}{\alpha-2} \Omega\left(\frac{jtP_{5}}{D^{\alpha}}\right)
\label{eq:proof:appendix:A:5}
\end{align}
\noindent where $\Omega(z) = {}_2F_1\left(1, 1 - \frac{2}{\alpha}; 2 - \frac{2}{\alpha}; z\right)$ is the hypergeometric function.

\noindent From \eqref{eq:proof:appendix:A:5} and \eqref{eq:proof:appendix:A:4}, the CF of the interferences is:
\begin{align}
\Phi_{I_5}\left( t \mid r_s \right) 
&= \exp\! \left( \frac{jt \eta  P_{5} \,\pi\lambda_5 \left(D^2 - r_s^2 \right)}{1 - jt  P_{5}} \right) \nonumber\\
&\times\exp \! \left( \! \frac{jt \eta  P_{5} \, 2 \pi \lambda_5 D^{2 - \alpha}}{\alpha - 2} \, \Omega\left(\!\frac{j t P_{5} }{D^\alpha} \!\right) \! \right)
\label{eq:proof:appendix:A:6}
\end{align}
\textbf{For $\boldsymbol{r_s} \geq \boldsymbol{D}$}, we proceed in the same manner.

Finally, taking the expectation with respect to $r_s$ finishes the proof.

\section{Proof of the CF of EMF exposure in a 5G $\beta$-GPP} \label{Appendix:B}
Following the proof of Theorem~1 in \cite{gontier2024joint}, we have:
{\footnotesize
\begin{align}
\Phi_{\mathcal{E}_5}(t) 
&= \sum_s^\infty \mathbb{E}_{r_s} \Bigg[\exp\left(jt P_{5}  |h_s|^2 \ell (Q_s)\right) \nonumber\\
& \times \exp \left( jt\sum_{k \neq s} P_{5} G_k |h_k|^2 \ell \left(Q_k \right) \epsilon_k \right) \mathbf{1}_{\mathcal{A}_s} \mathbf{1}_{\epsilon} \mid r_s^2 \Bigg] \nonumber \\
&= \beta_5 \! \sum_s^\infty \mathbb{E}_{r_s} \! \! \left[ \mathbf{1}_{\mathcal{A}_s} \! \mid \! r_s^2 \right] \mathbb{E}_{r_s} \! \! \left[ \Phi_{S_5} \! \left(t \! \mid \! r_s^2\right) \! \right] \mathbb{E}_{r_s} \! \! \left[ \Phi_{I_5} \! \left(t \! \mid \! r_s^2\right) \! \right]
\label{eq:proof:appendix:B:1}
\end{align}
}
\noindent The CFs of the desired signal and the interference are:
\begin{align}
\Phi_{S_5} (t \mid r_s^2) 
&= \frac{1}{1 - jt P_{5} \, \ell\left(\frac{\beta_5}{c_5} u \right)}\\
\Phi_{I_5}\!(t \mid r_s^2) &= \mathbb{E} \left[ \exp \left( jt \sum_{k \neq s} P_{5} G_k |h_k|^2 \ell(Q_k) \epsilon_k \right) \mid r_s^2 \right] \nonumber \\
&= \prod_{k \neq s} \mathbb{E}_{\Psi} \left[ 1 - \eta\beta_5 + \frac{\eta\beta_5}{1 - jt P_{5}  \ell(Q_k)} \mid r_s^2 \right] \nonumber \\
&\overset{(a)}{=} \prod_{k \neq s} \! \left(1 \! - \! \eta\beta_5 \! + \! \int_{u}^{\infty} \! \frac{\eta\beta_5 \, f_k(v)}{1 \!-\! jt P_{5}  \ell(\frac{\beta_5}{c_5}v)} \, dv \right)
\label{eq:proof:appendix:B:3}
\end{align}
where (a) follows from the change of variables in \eqref{eq:pdf:gpp}, and for the indicator term, we have:
{\small
\begin{align}
\mathbb{E}\left[ \mathbf{1}_{\mathcal{A}_s} \mid r_s^2 \right] &= \mathbb{E}_{\Psi_5} \left[ \prod_{k \neq s} \mathbb{E}_{\epsilon} \left[ \mathbf{1}_{ \left\{ \left\{ \epsilon_k = 1,\ r_k^2 > r_s^2 \right\} \cup \left\{ \epsilon_k = 0 \right\} \right\} } \right] \mid r_s^2 \right] \nonumber \\
&= \prod_{k \neq s} \left( 1 - \beta_5 + \beta_5 \, \mathbb{E}_{\Psi} \left[ \mathbf{1}_{ \{ r_k^2 > r_s^2 \} } \right] \mid r_s^2 \right) \nonumber \\
&= \prod_{k \neq s} \left( 1 - \beta_5 + \beta_5 \, \frac{\Gamma(k, u)}{\Gamma(k)} \right)
\label{eq:proof:appendix:B:2}
\end{align}
}


\section{Proof of the CDF of EMF exposure in 5G EN-DC} \label{Appendix:C}
The nearest site is always equipped with a 4G BS:
\begin{align}
\Phi_{\mathcal{E}}^\star(t) &= \mathbb{E} \left[ e^{jt \left( S_4 + S_5 \right)} \, e^{jt\left( I_4 + I_5 \right)} \right] \nonumber \\
&= \int_0^\infty \Phi_{S}^\star(t \mid \theta_4) \, \Phi_{I}^\star(t \mid \theta_4) \, d\Pi_4(\theta_4)
\label{eq:proof:appendix:C:1}
\end{align}

The nearest 4G BS serves the user. This BS may have a co-located 5G BS with probability $p$, which then serves as the 5G serving BS.
\begin{align}
\Phi_{S}^\star(t \! \mid \! \theta_4) \!&= \! \mathbb{E} \! \left[e^{jtP_4|h_{s,4}|^2\ell(\theta_4)} \! \mid \! \theta_4 \right]  \mathbb{E} \! \left[ e^{jt \zeta_s P_5 |h_{s,5}|^2 \ell(\theta_5)} \! \mid \! \theta_4\right] \nonumber \\
\!&=\! \frac{1}{1 \!-\! jtP_4\ell(\theta_4)} \left( 1 - p + \frac{p}{1 \!-\! jtP_5 \ell(\theta_4)} \right)
\label{eq:proof:appendix:C:2}
\end{align}

\noindent For the CF of the interference in $\beta$-GPP, we have:
\begin{align}
\Phi_{I}^\star(t \mid u) &= \prod_{k \neq s} \mathbb{E} \left[ \exp\left(jtP_4 \left|h_{k, 4} \right|^2 \ell \left(\frac{\beta_4}{c_4} v \right) \epsilon_k \right) \mid u \right] \nonumber \\
& \mathbb{E} \left[ \exp\left(jt \zeta_k P_5 G_{k,5} \left|h_{k, 5} \right|^2 \ell\left(\frac{\beta_4}{c_4} z \right) \right) \mid u, \epsilon_k = 1 \right] \nonumber \\
&= \prod_{k \neq s} \Bigg( 1 - \beta_4 + \int_u^\infty \frac{\beta_4}{1 - jt P_4 \ell\left( \frac{\beta_4}{c_4}v \right)} \nonumber\\
&\quad \left( 1\! - \!p \eta\! +\! \frac{p \eta}{1 \!-\! jt P_5 \ell\left( \frac{\beta_4}{c_4} v \right)} \right) f_k(v) \, dv \Bigg)
\label{eq:proof:appendix:C:3}
\end{align}

\noindent For PPP, by following the $\beta$-GPP approach and applying the PGFL, we obtain:
\begin{align}
\Phi_{I}^{\star}(t\mid r_4) 
&= \exp \Bigg( -2 \pi \lambda_4 \int_{r_4}^\infty \Bigg( 1 - \frac{1}{ 1 - jt P_4 \ell\left( r^2 \right)} \nonumber \\
& \qquad \left( 1 - p\eta + \frac{p\eta}{1 - jt P_5 \ell\left( r^2 \right) } \right) r \, dr \Bigg) \Bigg)
\label{eq:proof:appendix:C:4}
\end{align}
Substituting \eqref{eq:proof:appendix:C:2}, \eqref{eq:proof:appendix:C:3} and \eqref{eq:proof:appendix:C:4} into \eqref{eq:proof:appendix:C:1} completes the proof.

\section{Proof of the CDF of REBT-DL in 5G networks} \label{Appendix:D}
For every threshold $y>0$: $Y \, \triangleq \, \frac{S_5 + I_5}{W_5 \log_2 \left(1 + \frac{S_5}{I_5}\right)} \leq y$, and for
fixed $s_5 = h P_5 \ell(\theta)$, where channel gain and distance to the nearest BS are known:
\begin{align}
g(I_5;s_5) \, \triangleq \, s_5 + I_5 - yW_5 \log_2 \left(1+\frac{s_5}{I_5 }\right)
\label{eq:appendix:D:2}
\end{align}
Then $
{Y\le y} \quad\Longleftrightarrow \quad {g(I_5;s_5)\le 0}$.\\
We can see that, for each $S$: $\frac{\partial g(I_5;s_5)}{\partial I_5} =1 \, + \, \frac{y \, W_5}{\ln 2} \cdot \frac{s_5}{I_5(I_5 + s_5)} \, > \, 0$.
That means $g(\cdot;s_5)$ increases strictly on $I_5 \! \ge \! 0$, there is a unique root $I_5^\star$ of $g(I_5;s_5)\!=\! 0$. Therefore, the problem becomes $I_5 \! \leq \! I_5^\star$.
{\footnotesize
\begin{align*}
\mathbb{P} \left( Y < y \right) \nonumber 
&= \int_0^\infty \! \! \! \! \int_0^\infty F_{I} \left( I_5^\star \! \mid \! \theta \right) \, f_{|h|^2}(h) \, dh \, d\Pi(\theta) \nonumber \\
&= \int_0^\infty \! \! \! \! \int_0^\infty \! \! \left( \frac{1}{2} \!-\! \int_0^\infty \! \!\frac{\mathrm{Im}\{ \Phi_I^{(\text{5G})}(t \! \mid \! \theta) e^{-jtI_5^\star} \}}{\pi t} dt \! \right) e^{-h} dh \, d\Pi(\theta)
\end{align*}
}

\section{Proof of the CDF of REBT-DL in EN-DC network} \label{Appendix:E}
For $y>0$, REBT-DL in 5G EN-DC network is:
\begin{align}
Y \!=\! \frac{(S_4+I_4)+(S_5+I_5)}{W_4\log_2 \! \left(1+\frac{S_4}{I_4} \right) + W_5\log_2 \! \left(1 + \frac{S_5} {I_5} \right)} \le y
\label{eq:appendix:E:1}
\end{align}
Inequality \eqref{eq:appendix:E:1} can be rewritten as: $g(I_4;S_4) + g(I_5;S_5) \le 0$.
Fixing $s_5,i_4$ and $s_4$, the inequality becomes
$I_5\le I_5^\star$, where $I_5^\star$ is the unique solution of $g(i_4;s_4) + g(I_5;s_5) = 0$.

\noindent Taking the conditions on $ s_5 = \zeta_s P_5h_5 \ell(\theta_5),  s_4 = P_4 h_4 \ell(\theta_4) $, and $ i_4 $, and since these random variables are independent of each other and also independent of $ I_5 $, we have:
\begin{align}
&\mathbb{P} (Y < y) \nonumber\\
&= \mathbb{E}_\zeta \left[ \mathbb{E}_\Psi \left[ \mathbb{E}_{|h|^2} \left[ \mathbb{P} \left( I_5 < I_5^\star \! \mid \! i_4, h_4, h_5, \theta_4, \theta_5 \right) \right] \right] \right] \nonumber \\
&= \! \iiint \!\!\! \int_0^\infty \! \left( p \, \mathbb{P} \left(I_5 < I_{5, co}^\star \right) + (1-p) \mathbb{P} \left(I_5 < I_{5, no-co}^\star \right) \right) \nonumber\\
& \quad \quad \quad \quad f_{I_4}(i_4) \, e^{-h_4} \, e^{-h_5} \, d_{i_4} \, dh_4 \, dh_5 \, d\Pi_4(\theta_4)
\label{eq:appendix:E:3}
\end{align}
where $ \mathbb{P}\left(I_5 < I_{5, co}^\star\right) $ is the CDF of $ I_5 $ in the case where the 5G serving cell is co-located with the 4G serving cell, $ \ell(\theta_5) = \ell(\theta_4) $. In this case, the CF of the 5G interference is derived in the same way as in Proposition 2, but with the 4G signal component removed. 

\noindent When the serving 4G (located at $\theta_4$) and 5G BSs (located at $\theta_5 > \theta_4$) are positioned on different sites, we have:
\begin{align}
\mathbb{P} \!\left(\!I_5 \!<\! I_{5, no-co}^\star \!\right) \!=\! \int_{\theta_4}^\infty \! F_{I_5}\! \left( \! I_{5, no-co}^\star \!\mid \! \theta_5 \! \right) f_{\Theta_5|\Theta_4}\left( \! \theta_5 \!\mid \! \theta_4 \!\right)
\end{align}

{\itshape For PPP} ($\theta_4 = r_4 < \theta_5 = r_5$). The probability that no 5G BS exists within 2 circles with radius $r_4$ and $r_5$ is:
\begin{align}
\mathbb{P} \left( R_5 > r_5 \! \mid \! R_4 = r_4 \right) &= \mathbb{P} \left( \mathcal{N} \left\{ \pi \left( r_5^2 - r_4^2 \right) \right\} = 0 \right) \nonumber \\
&= \exp\left(-\pi p \lambda_4 \left(r_5^2 - r_4^2 \right)\right)
\label{eq:appendix:E:4}
\end{align}

{\itshape For $\beta$-GPP} ($\theta_4 = u < \theta_5 = v$). For each site $k \neq s$, the probability that no 5G BS is located between $u$ and $v$ is:
\begin{align}
\mathbb{P}\left( \text{no 5G} < v \mid \mathcal{A}_s, k\right) &= \frac{\mathbb{P}\left( \text{no 5G} < v, \mathcal{A}_s\right)}{\mathbb{P} \left( {\mathcal{A}}_s \right)}
\end{align}
The event no 5G inside $[u,v)$ at site $k$ while respecting the serving condition $\mathcal{A}_s$ is equivalent to the disjoint union: $\{\epsilon_k=0\}
\ \cup \{\epsilon_k=1,\ V_k>u,\ \zeta_k=0\}
\ \cup \{\epsilon_k=1,\ \zeta_k=1,\ V_k>v\}$.
Therefore, the probability that no 5G BS exists between $u$ and $v$ on all interfering sites is:
{\small
\begin{align}
\mathbb{P}\left( \text{no 5G} < v \mid \mathcal{A}_s\right)
= \prod_{k \neq s} \left( 1+\frac{\beta_4 p\left(\frac{\Gamma(k,v)}{\Gamma(k)}-\frac{\Gamma(k,u)}{\Gamma(k)}\right)}{1-\beta_4+\beta_4\frac{\Gamma(k,u)}{\Gamma(k)}}  \right)
\label{eq:appendix:E:5}
\end{align}
}
Then, taking the derivative with respect to $r_5$ and $v$ for the CCDF in \eqref{eq:appendix:E:4} and \eqref{eq:appendix:E:5}, we have the PDF.



\ifCLASSOPTIONcaptionsoff
  \newpage
\fi





\bibliographystyle{IEEEtran}
\bibliography{IEEEabrv,Bibliography}

\begin{thebibliography}{10}
\providecommand{\url}[1]{#1}
\csname url@rmstyle\endcsname
\providecommand{\newblock}{\relax}
\providecommand{\bibinfo}[2]{#2}
\providecommand\BIBentrySTDinterwordspacing{\spaceskip=0pt\relax}
\providecommand\BIBentryALTinterwordstretchfactor{4}
\providecommand\BIBentryALTinterwordspacing{\spaceskip=\fontdimen2\font plus
\BIBentryALTinterwordstretchfactor\fontdimen3\font minus
  \fontdimen4\font\relax}
\providecommand\BIBforeignlanguage[2]{{%
\expandafter\ifx\csname l@#1\endcsname\relax
\typeout{** WARNING: IEEEtran.bst: No hyphenation pattern has been}%
\typeout{** loaded for the language `#1'. Using the pattern for}%
\typeout{** the default language instead.}%
\else
\language=\csname l@#1\endcsname
\fi
#2}}
\renewcommand\BIBentryALTinterwordstretchfactor{4}

\bibitem{andrews2014will}
J.~G. Andrews, S.~Buzzi, W.~Choi, S.~V. Hanly, A.~Lozano, A.~C. Soong, and
  J.~C. Zhang, ``What will 5g be?'' \emph{IEEE Journal on selected areas in
  communications}, vol.~32, no.~6, pp. 1065--1082, 2014.

\bibitem{celaya20202g}
M.~Celaya-Echarri, L.~Azpilicueta, J.~Karpowicz, V.~Ramos, P.~Lopez-Iturri, and
  F.~Falcone, ``From 2g to 5g spatial modeling of personal rf-emf exposure
  within urban public trams,'' \emph{IEEE Access}, vol.~8, pp.
  100\,930--100\,947, 2020.

\bibitem{deng2014ginibre}
N.~Deng, W.~Zhou, and M.~Haenggi, ``The ginibre point process as a model for
  wireless networks with repulsion,'' \emph{IEEE Transactions on Wireless
  Communications}, vol.~14, no.~1, pp. 107--121, 2014.

\bibitem{gontier2024joint}
Q.~Gontier, C.~Wiame, S.~Wang, M.~Di~Renzo, J.~Wiart, F.~Horlin, C.~Tsigros,
  C.~Oestges, and P.~De~Doncker, ``Joint metrics for emf exposure and coverage
  in real-world homogeneous and inhomogeneous cellular networks,'' \emph{IEEE
  Transactions on Wireless Communications}, vol.~23, no.~10, pp.
  13\,267--13\,284, 2024.

\bibitem{al2020statistical}
M.~Al~Hajj, S.~Wang, L.~Thanh~Tu, S.~Azzi, and J.~Wiart, ``A statistical
  estimation of 5g massive mimo networks’ exposure using stochastic geometry
  in mmwave bands,'' \emph{Applied Sciences}, vol.~10, no.~23, p. 8753, 2020.

\bibitem{wang2019meta}
S.~Wang and M.~Di~Renzo, ``On the meta distribution in spatially correlated
  non-poisson cellular networks,'' \emph{EURASIP Journal on Wireless
  Communications and Networking}, vol. 2019, no.~1, p. 161, 2019.

\bibitem{liu2024assessment}
J.~Liu, Y.~Zhang, W.~B. Chikha, S.~Wang, T.~Samaras, O.~Jawad, L.~Ourak,
  E.~Conil, and J.~Wiart, ``Assessment of emf exposure induced by wireless
  cellular phones in various usage scenarios in france,'' \emph{IEEE Access},
  2024.

\bibitem{baccelli2024random}
F.~Baccelli, B.~B{\l}aszczyszyn, and M.~Karray, ``Random measures, point
  processes, and stochastic geometry,'' 2024.

\bibitem{gil1951note}
J.~Gil-Pelaez, ``Note on the inversion theorem,'' \emph{Biometrika}, vol.~38,
  no. 3-4, pp. 481--482, 1951.

\bibitem{ANFR_Cartoradio}
{Agence Nationale des Fréquences (ANFR)}, ``Cartoradio,'' Available:
  \url{https://www.cartoradio.fr/index.html}, 2025.

\bibitem{gontier2024modeling}
Q.~Gontier, C.~Tsigros, F.~Horlin, J.~Wiart, C.~Oestges, and P.~De~Doncker,
  ``Modeling the spatial distributions of macro base stations with homogeneous
  density: Theory and application to real networks,'' \emph{arXiv preprint
  arXiv:2409.05468}, 2024.

\bibitem{andrews2011tractable}
J.~G. Andrews, F.~Baccelli, and R.~K. Ganti, ``A tractable approach to coverage
  and rate in cellular networks,'' \emph{IEEE Transactions on communications},
  vol.~59, no.~11, pp. 3122--3134, 2011.

\end{thebibliography}

\vfill
\end{document}